\documentclass[
final
 , nomarks
]{dmtcs-episciences}

\usepackage[utf8]{inputenc}
\usepackage{subfigure}

\usepackage[round]{natbib}

\author{Khaled Elbassioni\affiliationmark{1}}
 
\title[Dualization over Distributive Lattices]{On Dualization over Distributive Lattices}
\affiliation{
 Khalifa University of Science and Technology, Abu Dhabi, UAE}
\received{2020-08-29}

\revised{2021-06-17, 2022-07-20}

\accepted{2022-09-22}

\usepackage{amsmath,amssymb,graphicx,amsfonts,epsf}
\usepackage[noend]{algpseudocode}
\usepackage{algorithm}
\usepackage{tikz,xcolor}
\usepackage{graphicx} 
\usetikzlibrary{backgrounds,decorations.pathmorphing,arrows}
\usepackage{amsopn}

\def\RR{\mathbb R}

\def\cA{\mathcal A}
\def\cB{\mathcal B}
\def\cC{\mathcal C}
\def\cD{\mathcal D}
\def\cF{\mathcal F}

\def\cH{\mathcal H}
\def\cI{\mathcal I}

\def\cP{\mathcal P}

\def\cH{\mathcal H}

\def\cX{\mathcal X}

\def\b1{\mathbf 1}

\def\cI{{\mathcal I}}
\def\cY{{\mathcal Y}}

\newcommand{\raf}[1]{(\ref{#1})}
\newcommand{\proof}{\noindent {\bf Proof}~~}

\newcommand{\poly}{\operatorname{poly}}

\newcommand{\supp}{\operatorname{supp}}

\newcommand{\hide}[1]{}

\def \DL{\textsc{Dual}}
\def \LDL{\textsc{Lattice-Dual}}
\def \SDL{\textsc{Simple-Dual}}
\def \Enum{\textsc{Enum}}
\def \EnumInc{\textsc{Enum-Inc}}
\def \EnumJoinInc{\textsc{Enum-Joint-Inc}}

\newtheorem{theorem}{Theorem}

\newtheorem{corollary}{Corollary}
\newtheorem{proposition}{Proposition}

\begin{document}
\publicationdetails{24}{2022}{2}{7}{6742}
\maketitle
\begin{abstract}
  Given a partially ordered set (poset) $P$, and a pair of families of ideals $\cI$ and filters $\cF$ in $P$ such that each pair $(I,F)\in \cI\times\cF$ has a non-empty intersection, the dualization problem over $P$ is to check whether there is an ideal $X$ in $P$ which intersects every member of $\cF$ and does not contain any member of $\cI$.  Equivalently, the problem is to check for a distributive lattice  $L=L(P)$, given by the poset $P$ of its set of joint-irreducibles, and two given antichains $\cA,\cB\subseteq L$ such that  no $a\in\cA$ is dominated by any $b\in\cB$, whether $\cA$ and $\cB$ cover (by domination) the entire lattice. We show that the problem can be solved in quasi-polynomial time in the sizes of $P$, $\cA$ and $\cB$, thus answering an open question in Babin and Kuznetsov (2017). As an application, we show that minimal infrequent closed sets of attributes in a relational database, with respect to a given implication base of maximum premise size of one, can be enumerated in incremental quasi-polynomial time. 
\end{abstract}

\section{Introduction}
\label{intro}

Let $P$ be  a partially order set (poset). Denote by ``$\preceq$" the partial order defined on the elements of $P$. A subset $X\subseteq P$ is called an {\it ideal} if it is ``down-closed", that is, $x\preceq y$ and $y\in X$  imply that $x\in X$. Likewise, $X\subseteq P$ is called a {\it filter} if it is ``up-closed", that is, $y\preceq x$ and $y\in X$  imply that $x\in X$. We denote respectively by $\cI(P)$ and $\cF(\cP)$ the families of ideals and filters of the poset $P$. Note that a set $X\subseteq P$ is an ideal if and only if its complement $\overline X:=P\setminus X$ is a filter.

\medskip

Let $\cI\subseteq\cI(P)$ and $\cF\subseteq\cF(P)$ be a pair of families satisfying
\begin{align}\label{dual-cond}
I\cap F\neq\emptyset\text{ for all } I\in\cI \text{ and }F\in\cF. 
\end{align}
We say that $(\cI,\cF)$ form a {\it dual} pair in $P$ if for every $X\in\cI(P)$, either $X\supseteq I$ for some $I\in\cI$ or $\overline X\supseteq F$ for some $F\in\cF$. Also, by definition, the pair $(\cI,\emptyset)$ (resp., $(\emptyset,\cF)$) is dual if and only if $\cI\supseteq \{\emptyset\}$ (resp., $\cF\supseteq \{\emptyset\}$); such pairs will be called ``trivial".

\medskip

In this paper, we are interested in the following problem:
\begin{itemize}
	\item[] \DL$(P,\cI,\cF)$: Given a poset $P$ and a pair of sets $\cI\subseteq\cI(P)$ and $\cF\subseteq\cF(\cP)$ satisfying~\raf{dual-cond}, either find an ideal $X\in\cI(P)$ such that
	\begin{align}\label{witness}
	X\not\supseteq I \text{ for all }I\in\cI\text{ and } X\cap F\ne\emptyset\text{ for all }F\in\cF, 
	\end{align}
	or declare that no such $X$ exists, that is, $(\cI,\cF)$ is a dual pair. 
\end{itemize}
See Figure~\ref{f1} for an example.
It is useful to note that condition \raf{witness} is {\it symmetric} in $\cI$ and $\cF$: 
$X\in\cI(P)$ satisfies \raf{witness} for the pair $(\cI,\cF)$
if and only if $\overline X\in\cF(P)$ satisfies \raf{witness} for $(\cF,\cI)$. Note also that, without the intersection condition~\ref{dual-cond}, the problem is already NP-hard \cite{GK99} even if $P$ is an antichain  (a set of incomparable elements).

\medskip

In the special case when $P$ is an {\it antichain}, problem \DL$(P,\cI,\cF)$ reduces to the well-known {\it hypergraph transversal} problem: given two hypergraphs $\cI,\cF\subseteq 2^P$  satisfying \raf{dual-cond} check whether  $\cF$ is  the {\it transversal hypergraph}\footnote{Given a hypergraph $\cH\subseteq 2^P$,  its transversal hypergraph is the hypergraph containing all minimal transversals (i.e., hitting sets) for $\cH$.} of $\cI$.  Fredman and Khachiyan \cite{FK96} gave a quasi-polynomial algorithm for solving this problem that runs in time $O(|P|)(|\cI|+\cF|)^{o(\log(|\cI|+\cF|))}$, thus providing evidence that the problem is unlikely to be NP-hard. \cite{BK17} considered the more general problem \DL$(P,\cI,\cF)$ for any given poset $P$, and showed that a simpler algorithm in \cite{FK96} can be generalized to give a subexponential algorithm for solving \DL$(P,\cI,\cF)$,  whose running time is $2^{O(|P|^{0.67}\log^3(|\cI|+\cF|))}$. They asked whether the problem can be solved in quasi-polynomial time. 
As explained below, we answer this question in the affirmative. 

\begin{theorem}\label{t-main}
	Problem \DL$(P,\cI,\cF)$ can be solved in time $(m+k)^{o(m+k)}\poly(n)$, where $n:=|P|$, $m:=|\cI|$ and $k:=|\cF|$. 
\end{theorem}

Thus, introducing the partial order on $P$ does not (essentially) present any additional difficulty. We prove Theorem~\ref{t-main} in the sections~\ref{decomp} and \ref{sec:alg}. Our algorithm is a generalization of the one given in \cite{E08} for the hypergraph transversal problem. In Section~\ref{lattices}, we state an equivalent formulation of Theorem~\ref{t-main} over distributed lattices and derive from it that dualization over products of distributive lattices can be done in quasi-polynomial time, thus improving on some of the results in \cite{E09}. As an application, we consider in Section~\ref{impl} the distributive lattice defined by the closed sets in an implication base of dimension one, and show that the minimal closed sets satisfying a given monotone system of transversal inequalities can be enumerated in incremental quasi-polynomial time. These include, for instance, minimal {\it infrequent} closed sets, i.e., those supported by small number of rows in the database. 

\section{Dualization over distributive lattices}\label{lattices}
A lattice $L$ is a poset in which every pair of elements $x,y\in L$ has a {\it greatest lower bound}\footnote{that is, an element $z\in L$ satisfying $w\preceq z$, for any $w\in L$ such that $w\preceq x$ and $w\preceq y$.} (meet), denoted by $x\wedge y$, 
and a {\it least upper bound}\footnote{that is, an element $z\in L$ satisfying $z\preceq w$, for any $w\in L$ such that $x\preceq w$ and $y\preceq w$.}  (join), denoted by $x\vee y$. 
Let $\cA,\cB\subseteq L$ be a pair of antichains in $L$ satisfying
\begin{align}\label{dual-cond1}
a\not\preceq b \text{ for all } a\in\cA \text{ and }b\in\cB. 
\end{align}
We say that $(\cA,\cB)$ is a {\it dual} pair in $L$ if for every $x\in L$, either $a\preceq x$ for some $a\in\cA$ or $x\preceq b$ for some $b\in\cB$. 

\medskip

The following problem has been considered in a number of papers (see, e.g., \cite{E09,BK17,DNU19,DN20}):
\begin{itemize}
	\item[] \LDL$(L,\cA,\cB)$: Given a lattice $L$ and a pair of antichains $\cA,\cB\subseteq L$ satisfying~\raf{dual-cond1}, either find an element $x\in L$ such that
	\begin{align}\label{witness1}
	a\not \preceq x \text{ for all }a\in\cA\text{ and } x\not\preceq b\text{ for all }b\in\cB, 
	\end{align}
	or declare that no such $x$ exists, that is, $(\cA,\cB)$ is a dual pair.  
\end{itemize}
When $L$ is given as a Cartesian product of $n$ lattices $L:=L_1\times\cdots\times L_n$, it was shown in \cite{E09} that the problem can be solved in time $\poly(\sum_{i=1}^n|L_i|)(|\cA|+\cB|)^{o(\log(|\cA|+\cB|)W^2\log W)}$, where $W$ is the maximum size of an antichain in any of the $L_i$'s. 

\begin{figure}
	\centering
	\includegraphics[width=3.5in]{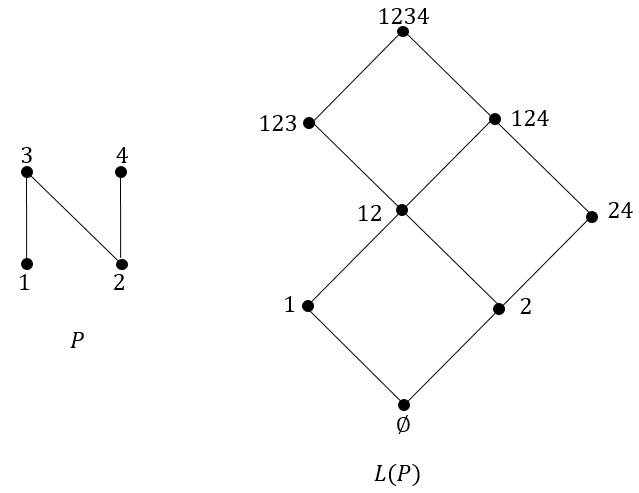}
	\caption{A poset $P=\{1,2,3,4\}$ and the corresponding lattice $L=L(P)$; edges from an element $x$ to a higher (in the drawing) element $y$ indicate the order $x\preceq y$. The set of joint-irreducibles of $L$ is $J(L):=\{1,2,24,123\}$.
		The families of ideals $\cI=\big\{\{1,2\},\{2,4\}\big\}$ and filters $\cF=\big\{\{2,3,4\}\big\}$ form a dual pair $(\cI,\cF)$ in $P$; the corresponding dual pair in the lattice $L(P)$ is $(\cA,\cB)$, where $\cA=\{12,24\}$ and $\cB=\{1\}$.
	}
	\label{f1}
\end{figure}

Denote by $0_{L}:=\bigwedge_{x\in L_i}x$ the meet of all elements in $L$. An element $x\in L$ such that $ x\ne0_L$, is said to be {\it join-irreducible} if $x$  cannot be written as the join of any two distinct elements of $L$. Denote by $J(L)$ the poset of join-irreducible elements of $L$. A lattice in which the operations of join and meet distribute over each other is said to be {\it distributive}. 
The well-known {\it Birkhoff's representation theorem} (see, e.g., \cite[Chapter 3]{SF99}) states that any (finite) distributive lattice $L$ can be represented as the lattice $L(P)$ of ideals of the poset $P$ defined (by the lattice order) on $J(L)$, ordered by set-inclusion. See the example in Figure~\ref{f1}. In particular, any element $x\in L$ has a unique representation as: 
\begin{align}\label{JI}
x=\bigvee_{y\in J(L):~y\preceq x}y.
\end{align}
In view of this theorem, we obtain immediately the following result as a corollary of Theorem~\ref{t-main}.

\begin{corollary}\label{c1}
	For any distributive lattice $L$, given by its poset of join-irreducibles $J(L)$, and antichains $\cA,\cB\subseteq L$ given by their representation~\raf{JI}, problem \LDL$(L,\cA,\cB)$ can be solved in time $(m+k)^{O(\chi(m+k))}\poly(n)$, where $n:=|J(L)|$, $m:=|\cA|$ and $k:=|\cB|$.
\end{corollary}

Let $L:=L_1\times\cdots\times L_n$ be the product of $n$ distributive lattices $L_1,\ldots,L_n$, each given by an explicit list of its elements. Then 
\begin{align}\label{JI2}
J(L)=\bigcup_{i=1}^n J_i,\text{ where } J_i:=\{(0_{L_{1}},\ldots,0_{L_{i-1}},p,0_{L_{i+1}},\ldots,0_{L_{n}}):~p\in J(L_i)\}.
\end{align}
It follows from \raf{JI2} that $|J(L)|\le\sum_{i=1}^n|J(L_i)|$. Thus, we obtain the following result as a consequence of Corollary~\ref{c1}.

\begin{corollary}\label{c2}
	Problem \LDL$(L,\cA,\cB)$ can be solved in time $(m+k)^{O(\chi(m+k))}\poly(n)$, if $L:=L_1\times\cdots\times L_n$ is the product of $n$ distributive lattices, where $n:=\sum_{i=1}^n|L_i|$, $m:=|\cA|$ and $k:=|\cB|$.
\end{corollary}

Corollary~\ref{c2} improves on the result in \cite{E09} when each $L_i$ is a distributive lattice\footnote{Indeed, even though the algorithm in \cite{E09} has a quasi-polynomial time complexity in terms of $m$ and $k$ for the product of {\it general} lattices, it has {\it exponential} complexity in terms of the maximum width of the lattices $L_1,\ldots,L_n$.}.
When each $L_i$ is chain, a stronger bound on the running time, that does on the size of each chain, was given in \cite{BEGKM02-SICOMP}.

\section{Decomposition rules}\label{decomp}

Let $n:=|P|$, $m:=|\cI|$ and $k:=|\cF|$. For a set $S\subseteq P$ and a family $\cH\subseteq 2^P$, let $\cH_S:=\{H\in\cH:~H\subseteq S\}$ be the subfamily of $\cH$ induced by set $S$, $\cH^S=\{H\cap S:~H\in\cH\}$ be the projection (or trace) of $\cH$ into $S$ (this can be a multi-subfamily), and $\cH(S):=\{H\in\cH:~H\cap S\ne\emptyset\}$ be the subfamily with sets having non-empty intersection with $S$. We define further $\overline S:=P\setminus S$,
and for $p\in P$, let $\deg_{\cH}(p):=|\{H\in\cH:~p\in H\}|$ be the degree of $p$ in $\cH$. 

For an element $p\in P$, we denote respectively by $p^+:=\{x\in P:~p\preceq x\}$ and $p^-:=\{x\in P:~x\preceq p\}$ the smallest ideal and filter containing $p$. Similarly, for $X\subseteq P$, we denote by $X^-:=\cup_{p\in X}p^-$ and $X^+:=\cup_{p\in X}p^+$ the smallest ideal and filter containing $X$, respectively. 

\medskip

\begin{figure}
	\centering
	\includegraphics[width=3.5in]{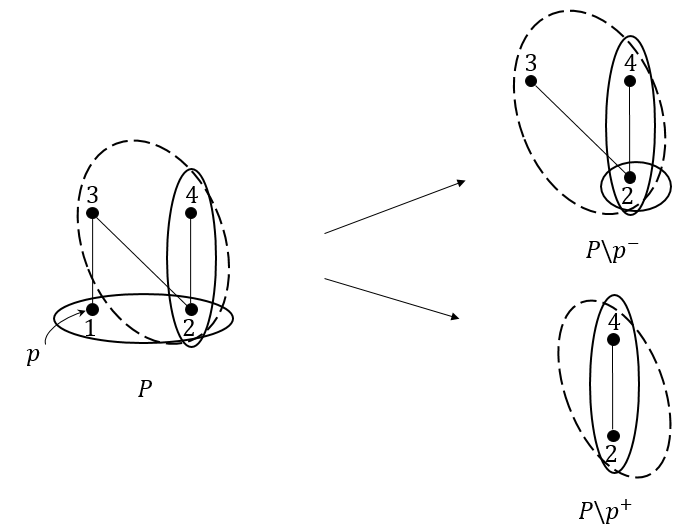}
	\caption{A poset $P=\{1,2,3,4\}$ and families of ideals $\cI=\big\{\{1,2\},\{2,4\}\big\}$ and filters $\cF=\big\{\{2,3,4\}\big\}$, shown by solid and dotted ellipses, respectively. The a decomposition into two subproblems based on Proposition~\ref{p1} is shown on the right, where $p=1$.  The corresponding families of ideals and filters in the two subproblems are
		$\cI'=\big\{\{2\},\{2,4\}\big\}$, $\cF'=\big\{\{2,3,4\}\big\}$ for the subproblem on $P\setminus p^-$, and	$\cI''=\big\{\{2,4\}\big\}$, $\cF''=\big\{\{2,4\}\big\}$ for the subproblem on $P\setminus p^+$.
	}
	\label{f3}
\end{figure}

The following proposition says that we can decompose a given duality testing problem  \DL$(P,\cI,\cF)$ into two duality testing subproblems, by choosing an arbitrary element $p\in P$ and  considering the two posets obtained by deleting all the elements dominating $p$, resp., dominated by $p$. The possible reductions in size of the resulting subproblems comes from the fact that in each subproblem we only need to consider either the induced subfamily of ideals and the projected subfamily of filters or vice versa (see Figure~\ref{f3} for an illustration). 
\begin{proposition}[\cite{BK17}]\label{p1}
	Let $\cI\subseteq \cI(P)$ and $\cF\subseteq\cF(P)$ be a non-trivial pair of families (satisfying~\raf{dual-cond}), $p\in P$ be a given
	element. Then the pair $(\cI,\cF)$ is dual in $P$ if and only if 
	\begin{align}\label{d0.1}
	(\cI^{\overline{p^-}},\cF_{\overline{p^-}})\text{ is dual in }P\setminus p^-,\text{ and }\\
	(\cI_{\overline{p^+}},\cF^{\overline{p^+}})\text{ is dual in }P\setminus p^+\label{d0.2}.
	\end{align}
\end{proposition}
\proof
For convenience of notation, let us write
\begin{align*}
\cI':=\cI^{\overline{p^-}}=\{I\setminus p^-:~I\in\cI\}, \qquad \cF':=\cF_{\overline{p^-}}=\{F\in\cF:~F\cap p^-=\emptyset\},\\
\cI'':=\cI_{\overline{p^+}}=\{I\in\cI:~I\cap p^+=\emptyset\}, \qquad \cF'':=\cF^{\overline{p^+}}=\{~F\setminus p^+:~F\in\cF\}.
\end{align*}
Assume that $(\cI',\cF')$ and $(\cI'',\cF'')$ are dual in $P\setminus p^-$ and $P\setminus p^+$, respectively. We show that $(\cI,\cF)$ is dual in $P$. Consider any $X\in\cI(P)$. Suppose that $p\in X$ (and hence, $p^-\subseteq X$). Then, by the duality of $(\cI',\cF')$, either $X\setminus p^-\supseteq I\setminus p^-$ for some $I\in\cI$, or $\big(P\setminus p^-\big)\setminus\big(X\setminus p^-\big)\supseteq F$ for some $F\in\cF$ such that $F\cap p^-=\emptyset$. In the former case,   $X\supseteq I$ follows by $X\supseteq p^-$, and in the latter case, $P\setminus X\supseteq F$.
The case when $p\not\in X$ can be treated symmetrically (by replacing ``ideal" by ``filter", $\cI$ by $\cF$, and $X$ by $P\setminus X$).
Thus, we conclude that the pair $(\cI,\cF)$ is dual in $P$.	

Conversely, assume that $(\cI,\cF)$ is dual in $P$. We first show that $(\cI',\cF')$ is dual in $P\setminus p^-$. Consider any $X\in\cI(P\setminus p^-)$. Then, by the duality of $(\cI,\cF)$, either $X\cup p^-\supseteq I$ for some $I\in\cI$, or  $P\setminus\big(X\cup p^-\big)\supseteq F$ for some $F\in\cF$. In the former case, we have $X\supseteq I\setminus p^-\in\cI'$, and in the latter case, we have $F\cap p^-=\emptyset$, and hence, $\big(P\setminus p^-\big)\setminus X\supseteq F\in\cF'$. Thus, we get that $(\cI',\cF')$ are dual in $P\setminus p^-$. A symmetric statement also shows that $(\cI'',\cF'')$ is dual in $P\setminus p^+$.
\qed

Note that both pairs in \raf{d0.1} and \raf{d0.2} satisfy the intersection condition~\raf{dual-cond}, whenever the original pair ($\cI,\cF)$ does.
Based on Proposition~\ref{p1}, a recursive algorithm with sub-exponential running time $2^{O(n^{0.67}\log^3(m+k))}$ was given in \cite{BK17}. The idea, following \cite{FK96}, was to show that, under the duality of $(\cI,\cF)$, there exists a {\it frequent} element in $P$, that is, a $p\in P$ such that either $\deg_{\cI}(p)\ge \epsilon|\cI|$ or $\deg_{\cF}(p)\ge \epsilon|\cF|$, for some $\epsilon\in(0,1)$. The currently best estimate of $\epsilon\ge\frac{1}{r^2\log_{4/3}(m+k)}$ that was shown in \cite{BK17}, where $r=\max_{p\in P}(|p^+|+|p^-|)$, implies the above mentioned subexponential time. In  order to improve on this, we need further decomposition ideas, which are provided by the next set of propositions.

\medskip

For a family of subsets $\cH\subseteq 2^\cP$ and a positive number $\epsilon\in(0,1)$, denote by $L(\cH,\epsilon)$ the subset $\{p\in P:~\deg_{\cH}(p)\ge \epsilon|\cH|\}$ of "large" degree vertices in $\cH$.
Given $\epsilon',\epsilon''\in(0,1)$, $\epsilon'<\epsilon''$, let us call an {\it $(\epsilon',\epsilon'')$-balanced set} with respect to $\cH$, any set $S\subseteq P$ such that $\epsilon'|\cH|\le|\cH_S|< \epsilon''|\cH|$.

The following proposition guarantees that if the set of ``large"-degree elements w.r.t. $\cI$ does not contain many ideals from $\cI$ (in fact, we will apply this where ``many" is replaced by ``any"), then we can find an ideal $S$ (not necessarily from $\cI$) that both contains and does not contain ``good" fractions of the ideals in $\cI$ (more precisely, such that the number of ideals in $\cI$ that are not contained in $S$ is in the interval $\big[\epsilon|\cI|,2\epsilon|\cI|\big]$, for a chosen $\epsilon$).  

\begin{proposition}[see \cite{E08}]\label{p2}
	Let $\cI\subseteq\cI(P)$ be a given set of ideals of $P$, and $\epsilon_1,\epsilon_2\in(0,1)$ be two given numbers such that, $\epsilon_1<\epsilon_2$ and $L=L(\cI,\epsilon_1)$ satisfies $|\cI_{L}|\leq(1-\epsilon_2)|\cI|$. Then there exists a $(1-\epsilon_2,1-(\epsilon_2-\epsilon_1))$-balanced ideal $S\supseteq L$ with respect to $\cI$. Such a set $S$ can be found in $O(nm)$ time, where $n=|P|$ and $m=|\cI|$.
\end{proposition}
\proof
Let $\{p_1,\ldots,p_l\}$ be 
an arbitrary order of the elements of $\overline L:=P\setminus L$ and $j\in[l-1]$ be the index 
such that 
\begin{equation}\label{part}
|\cI_{P\smallsetminus\{p_1,\ldots,p_j\}}|>(1-\epsilon_2) |\cI|\mbox{ and }
|\cI_{P\smallsetminus\{p_1,\ldots,p_{j+1}\}}|\le(1-\epsilon_2) |\cI|.
\end{equation}
The existence of such $j$ is guaranteed by the facts that $\deg_{\cI}(p_1)<\epsilon_1|\cI|< \epsilon_2|\cI|\leq|\cI(\overline L)|$. 
Finally, we let $S':=(P\smallsetminus\{p_1,\ldots,p_j\})$ and $S:=\bigcup_{I\in\cI_{S'}}I$. Since $\deg_{\cI}(p_{j+1})\le\epsilon_1|\cI|$, it follows from \raf{part} that $|\cI_{S}|<(\epsilon_1+1-\epsilon_2)|\cI|$, implying that $S$ is indeed a balanced superset of $L$. Note by definition that $S$ is an ideal.
\qed

By symmetry, an analogous version of Proposition~\ref{p2} can be stated as follows.

\begin{proposition}\label{p3}
	Let $\cF\subseteq\cF(P)$ be a given set of filters of $P$, and $\epsilon_1,\epsilon_2\in(0,1)$ be two given numbers such that, $\epsilon_1<\epsilon_2$ and $L=L(\cF,\epsilon_1)$ satisfies $|\cF_{L}|\leq(1-\epsilon_2)|\cF|$. Then there exists a $(1-\epsilon_2,1-(\epsilon_2-\epsilon_1))$-balanced filter $S\supseteq L$ with respect to $\cF$. Such a set $S$ can be found in $O(nk)$ time, where $n=|P|$ and $k=|\cF|$.
\end{proposition}

\medskip

Given the families $\cI$ and $\cF$, if there is a large-degree element $p\in P$ w.r.t. both families, then it can be used to decompose the problem according to Proposition~\ref{p1} with a  sufficient reduction in size in both of the resulting subproblems. On the other hand, if no such element exists, then by the intersection condition~\raf{dual-cond} and Propositions~\ref{p2} and~\ref{p3}, there must exist a balanced set $S$ w.r.t. either $\cI$ or $\cF$.  We can use this set $S$ for decomposing the duality testing problem in the following way. We first test duality ``inside" $S$; if there is a ``local" witness for non-duality inside $S$ (that is, there is an $X$ satisfying \raf{witness} w.r.t. the pair $\big(\cI_{S},\cF^{S}\big)$), it can be extended to a ``global" witness for non-duality in $P$ (that is, an  $X$ satisfying \raf{witness} w.r.t. the pair $\big(\cI,\cF\big)$), and we are done. Otherwise, the existence of a global witness $X$ but no local witness inside $S$ implies that there must exit a filter $F\in\cF$ whose projection $Y$ on $S$ is disjoint from $X$. Thus it is enough in this case to look for the witness $X$ inside the poset obtained from $P$ by deleting the subposet dominating $Y$. The following proposition describes such decomposition more formally.

\medskip

\begin{proposition}\label{p4}
	Let $\cI\subseteq \cI(P)$ and $\cF\subseteq\cF(P)$ be two a non-trivial pair of families (satisfying
	\raf{dual-cond}), and $S\in\cI(P)$ be a given
	ideal. Then $(\cI,\cF)$ is dual in $P$ if and only if
	\begin{eqnarray}
	\label{d1.1}
	&&\big(\cI_{S},\cF^{S}\big)\text{ is dual in } S,\text{ and }\\
	&&\big(\cI_{\overline {Y^+}},\cF^{\overline {Y^+}}\big) \mbox{ is dual in $\overline {Y^+}$, for all $Y\in\cF^S$}.  
	\label{d1.2}
	\end{eqnarray} 
\end{proposition}
\proof
Suppose that the pair $(\cI,\cF)$ is dual in $P$. If \raf{d1.1} is not satisfied, then 
there exists an $X\in\cI(S)$ such that $(S\setminus X)\cap I\neq\emptyset$ for all $I\in\cI_S$ and $X\cap F\neq\emptyset$ for all $F\in\cF^S$. Note that this implies, in  particular, that 
$\overline S\not\supseteq F$ for all $F\in\cF$, since we assume that $\cF\ne\emptyset$. Thus, $X\cap F\ne\emptyset$ for all $F\in\cF$. Note  also that $\overline X\cap I=((S\setminus X)\cup\overline S)\cap I\ne\emptyset$ for all $I\in\cI$. Thus $X\in\cI(P)$ satisfies~\raf{witness} for the pair $(\cI,\cF)$, in contradiction to its assumed duality. Replacing $S$ in the above argument by $\overline{Y^+}\in\cI(P)$, for $Y\in\cF^S$, shows that~\raf{d1.2} also holds.

Conversely, suppose that $(\cI,\cF)$ is not a dual pair, i.e., there exists an 
$X\in\cI(P)$ satisfying~\raf{witness}. If $X\cap S\in\cI(S)$ satisfies \raf{witness} for  the pair $(\cI_S,\cF^S)$, then~\raf{d1.1} is not satisfied. Otherwise, $S\setminus X \supseteq Y$ for some $Y\in\cF^S$, and consequently, $P\setminus X\supseteq Y^+$. In particular, $F\not\subseteq Y^+$ for any $F\in\cF$. Thus, if $\cI_{\overline{Y^+}}=\emptyset$, then $\big(\cI_{\overline {Y^+}},\cF^{\overline {Y^+}}\big)$ is (trivially) not dual. Otherwise, as $X\cap F \ne\emptyset$ for all $F\in\cF$, we must have $X\cap \big(F\setminus Y^+\big)\ne\emptyset$, for all $F\in\cF$. Also $\overline X\cap I\ne\emptyset$ for all $I\in\cI\supseteq\cI_{\overline{Y^+}}$. 
Thus, $X\in\cI(\overline{Y^+})$ is an evidence  that \raf{d1.2} is not satisfied.
\qed

By symmetry, an analogous version of Proposition~\ref{p4} can be stated as follows.

\medskip

\begin{proposition}\label{p5}
	Let $\cI\subseteq \cI(P)$ and $\cF\subseteq\cF(P)$ be a non-trivial pair of families (satisfying
	\raf{dual-cond}), and $S\in\cF(P)$ be a given
	filter. Then $(\cI,\cF)$ is dual in $P$ if and only if
	\begin{eqnarray}
	\label{d2.1}
	&&\big(\cI^{S},\cF_{S}\big)\text{ is dual in } S,\text{ and }\\
	&&\big(\cI^{\overline {Y^-}},\cF_{\overline {Y^-}})\mbox{ is dual in $\overline {Y^-}$, for all $Y\in\cI^S$}.  
	\label{d2.2}
	\end{eqnarray} 
\end{proposition}

\medskip

Observe that all pairs of $(\cI,\cF)$ arising in~\raf{d1.1},~\raf{d1.2}, ~\raf{d2.1} and~\raf{d2.2} satisfy condition \raf{dual-cond}, whenever the original pair ($\cI,\cF)$ does. 
Based on the above propositions, we can prove Theorem~\ref{t-main} using the algorithm given in the following section.

\section{A quasi-polynomial algorithm}\label{sec:alg}

Based on the decomposition rules described in Propositions~\ref{p1},~\ref{p4} and ~\ref{p5}, we give a recursive procedure that solves \DL$(P,\cI,\cF)$ as described in Algorithm~\ref{alg}. The procedure returns True or False depending on the duality/non-duality of the pair $(\cI,\cF)$. It can be easily modified to return a witness $X\subseteq \cI(P)$ in the case of non-duality. If $\min\{|\cI|,|\cF|\}\le 1$, then duality of $(\cI,\cF)$ can be easily tested using a simple procedure which we call \SDL.  (For instance, if $\cI=\{I\}$ for some $I\in\cI(P)$, we solve $|I|$ trivial subproblems, $\DL(P\setminus p^+,\emptyset,\cF^{P\setminus p^+})$, for all $p\in I$; the answer to each subproblem is True if and only if there is an $F\in\cF $ such that $F\subseteq p^+$).

For $v\in\RR_+$, let $\chi(v)$ be the unique positive root of the equation 
$$\left(\frac{\chi(v)}{2}\right)^{\chi(v)}=v.$$ 
Note that $\chi(v)\sim\log v/\log\log v$.

\begin{algorithm}[!htb]
	\caption{ \DL$(P,\cI,\cF)$} \label{alg}
	\begin{algorithmic}[1]
		\Require A poset $P$ and two families $\cI\subseteq\cI$ and $\cF\subseteq \cF(P)$ satisfying \raf{dual-cond} 
		\Ensure  True if $(\cI,\cF)$ is a dual pair and False otherwise 
		\If {$\min\{|\cI|,|\cF|\}\le 1$}
		\Return SIMPLE-\DL$(P,\cI,\cF)$\label{s1}
		\EndIf
		\State$v:=|\cI||\cF|$, $\epsilon_1:=\frac{1}{\chi(v)}$, $\epsilon_2:=\frac{2}{\chi(v)}$, $L_1:= L(\cI,\epsilon_1)$ and $L_2:= L(\cF,\epsilon_1)$
		\If {$|\cI_{L_1}|\geq 1$ and $|\cF_{L_2}|\ge 1$}\label{s3}
		/* by \raf{dual-cond}, $L_1\cap L_2\neq\emptyset$ */
		\State $p:=$an arbitrary element in $L_1\cap L_2$
		\State $d_1:=$\DL$(\overline{ p^-},\cI^{\overline{p^-}},\cF_{\overline{p^-}})$
		\State $d_2:=$\DL$(\overline{ p^+},\cI_{\overline{p^+}},\cF^{\overline{p^+}})$
		\EndIf
		\Return $d_1\wedge d_2$ 	/* cf. Proposition \ref{p1} */
		\If{$|\cI_{L_1}|=0$}\label{s4}
		\State $S_1:=$a $(1-\epsilon_2,1-(\epsilon_2-\epsilon_1))$-balanced superset of $L_1$
		/* cf. Proposition \ref{p2} */
		\State $d_1:=$\DL$(S_1,\cI_{S_1},\cF^{S_1})$
		\State $d_Y:=$\DL$(\overline {Y^+},\cI_{\overline {Y^+}},\cF^{\overline {Y^+}})$, for all $Y\in\cF^{S_1}$\\
		\Return $d_1\wedge(\bigwedge_{Y\in\cF^{S_1}} d_Y)$   /* cf. Proposition \ref{p4} */
		\Else /* $|\cF_{L_2}|=0$ */  \label{s5}
		\State $S_2:=$a $(1-\epsilon_2,1-(\epsilon_2-\epsilon_1))$-balanced superset of $L_2$   /* cf. Proposition \ref{p3} */
		\State $d_1:=$DUAL-P1$(S_2,\cI^{S_2},\cF_{S_2})$
		\State $d_Y:=$DUAL-P1$(\overline {Y^-},\cI^{\overline {Y^-}},\cF_{\overline {Y^-}})$, for all $Y\in\cI^{S_2}$\\
		\Return $d_1\wedge(\bigwedge_{Y\in\cI^{S_2}} d_Y)$   /* cf. Proposition \ref{p5} */
		\EndIf
	\end{algorithmic}
\end{algorithm}

\bigskip

\noindent{\bf Analysis of running  time.}~ Let  $T(v)$ be the number of recursive calls  
made by the algorithm on an instance of the problem of ``volume" $v=:|\cI||\cF|$.   

\medskip

\noindent{\em Step \ref{s3}:} Since $p\in L_1\cap L_2$, we have $\deg_{\cI}(p)\ge\epsilon_1|\cI|$ and $\deg_{\cF}(p)\ge\epsilon_1|\cF|$.
Thus $|\cI_{P\setminus p^+}|\le (1-\epsilon_1)|\cI|$, and $|\cF_{P\setminus p^-}|\le (1-\epsilon_1)|\cF|$, and we get the recurrence
\begin{eqnarray}\label{rec2-1}
T(v)&\le& 1+2T((1-\epsilon_1)v).
\end{eqnarray}

\medskip

\noindent{\em Step \ref{s4}:} Let $\epsilon:=\frac{|\cI(\overline S_1)|}{|\cI|}$. Then $|\cI_{S_1}|=(1-\epsilon)|\cI|$ and 
$|\cI_{\overline{Y^+}}|\le\epsilon|\cI|$ (since $\cI_{\overline{Y^+}}\subseteq\cI(\overline S_1)$ by \raf{dual-cond}) for all $Y\in\cF^{S_1}$. Since $|\cF^{S_1}|\leq |\cF| \leq v/2$ for $|\cI|> 1$, we get the recurrence
\begin{eqnarray}\label{rec2-2}
T(v)&\le& 1+T((1-\epsilon)v)+\frac{v}{2} T(\epsilon v),
\end{eqnarray} 
Furthermore, we know that
$\epsilon \in(\epsilon_2-\epsilon_1,\epsilon_2]$, since $S_1$ is a $(1-\epsilon_2,1-(\epsilon_2-\epsilon_1))$-balanced superset of $L_1$. 

\medskip

\noindent{\em Step \ref{s5}:} By symmetry we get the same recurrence as in \raf{rec2-2}.

\medskip

It can be shown (see \cite{FK96,E08}) by induction on $v\geq 0$ that the above recurrences imply that $T(v)\le v^{\chi(v)}$. We give the proof for completeness. For $v<4$, $T(v)=1$ follows from step \ref{s1} of the procedure. Now assume that $v\ge 4$. 
Consider \raf{rec2-1} . If $(1-\epsilon_1)v<4$, then the recurrence gives $T(v)\le 3<v^{\chi(v)}$, since $\chi(v)> 3$ for $v\geq 4$.
Otherwise, by induction and monotonicity of $\chi(\cdot)$, we get from \raf{rec2-1}
\begin{eqnarray*}
	T(v)&\le& 1+2((1-\epsilon_1)v)^{\chi(v)}\le 1+2\left(1-\frac{1}{\chi(v)}\right)^{\chi(v)}v^{\chi(v)}\\
	&\le& 1+\frac{2}{e}v^{\chi(v)}\le v^{\chi(v)},
\end{eqnarray*}
(where we used the inequality $1-x \leq e^{-x}$, valid for all real $x$) for $v$ large enough ($v\ge 4$ is sufficient, explaining our stopping criterion in step 1 of the algorithm).

Finally, consider recurrence \raf{rec2-2}. Assume that $(1-\epsilon)v\geq 4$ and $\epsilon v\geq 4.$ Then we get by induction
\begin{eqnarray*}
	T(v)&\le& 1+T\left(\left(1-\frac{1}{\chi(v)}\right)v\right)+\frac{v}{2}T\left(
	\frac{2v}{\chi(v)}\right) \\
	&\le& 1+\left[\left(1-\frac{1}{\chi(v)}\right)^{\chi(v)}+\frac{v}{2}\left(\frac{2}{\chi(v)}\right)^{\chi(v)}\right]v^{\chi(v)}\\
	&\leq& 1+\left[\frac{1}{e}+\frac{1}{2}\right]v^{\chi(v)}<v^{\chi(v)}.
\end{eqnarray*}
One can also verify that the inequality holds if $(1-\epsilon)v<4$ and/or $\epsilon v<4.$  \qed

\section{Enumerating minimal closed sets in an implication base satisfying a monotone property}\label{impl}

Let $\cD$ be a relational database defined on a finite set $P$ of {\it binary attributes} (or items). We think of any row in $\cD$ as a subset of $P$. An (unary) {\it implication base} $\Sigma$ over $P$ is a set of implications of the form $A\rightarrow  b$ where $A \subseteq P$ and $b\in P$ (see, e.g., \cite{W17,BDVG18}). In such implication, $A$ is called the {\it premise}\footnote{Naturally, one assumes that the base is {\it irredundant} in the sense that no premise contains another, as such containment would lead to a redundant implication.}   and $b$ is called the {\it conclusion}. The {\it dimension} of $\Sigma$ is the size of a largest premise: $\dim(\Sigma):=\max\{|A|: (A\rightarrow b)\in\Sigma\}$. A set $X \subseteq P$ is said to be {\it closed} w.r.t.  $\Sigma$ if $(A\rightarrow b)\in\Sigma$ and $A\subseteq X$ imply that $b\in X$. The {\it support} of $X$ in $\cD$, denoted by $\supp_\cD(X):=\{D\in\cD:~D\subseteq X\}$, is the set of rows in the database $\cD$ that contain $X$. Let us denote by $\cC(\Sigma)$ the family of all closed sets w.r.t. $\Sigma$. It is well-known (see, e.g., \cite{BDVG18}) that $\cC(\Sigma)$, ordered by set-inclusion, forms a lattice, and that every lattice arises this way from some $\Sigma$. When $\dim(\Sigma)=1$ then the lattice is distributive, and $\cC(\Sigma)=\cI(P)$, where the order on $P$ is defined by setting $b\preceq a$ whenever $(\{a\}\rightarrow b)\in\Sigma$. We are interested in {\it monotone enumeration} problems over $\cC(\Sigma)$:
\begin{itemize}
	\item[] \Enum$(P,\Sigma,\pi)$: Given an implication base $\Sigma$ over a  set $P$ and a monotone\footnote{that is, $\pi(X)\le \pi(Y)$ whenever $X\subseteq Y\subseteq P$.} property $\pi:2^P\to\{0,1\}$, enumerate all (inclusion-wise) minimal sets in $\cC(\Sigma)$ satisfying $\pi$. 
\end{itemize}

Let us denote by $\cC_{\pi}(\Sigma)$ the family of all minimal sets in $\cC(\Sigma)$ satisfying $\pi$.
Since the size of the output in such problems can be exponential in the size of the input, it is customary to measure the running time of an algorithm for solving them as a function of both the input and output sizes. In particular, an algorithm for enumerating the elements of the family $\cC_{\pi}(\Sigma)$ is said to run  in {\it output polynomial time} if it enumerates all the elements of $\cC_{\pi}(\Sigma)$ in time $\poly(|P|,|\Sigma|,|\pi|,|\cC_{\pi}(\Sigma)|)$, where $|\pi|$ is the size of the {\it encoding description} of $\pi$. To have a more refined definition of enumeration efficiency, we consider the following {\it incremental version} of  \Enum$(P,\Sigma,\pi)$: 
\begin{itemize}
	\item[] \EnumInc$(P,\Sigma,\pi,\cX)$: Given an implication base $\Sigma$ over a  set $P$, a monotone property $\pi:2^P\to\{0,1\}$, and a set $\cX\subseteq \cC_{\pi}(\Sigma)$, either find a set $X\in\cC_{\pi}(\Sigma)\setminus\cX$, or declare that no such set exists. 
\end{itemize}
We say that an algorithm $\cA$ for solving \Enum$(P,\Sigma,\pi)$ is {\it incremental polynomial time} if it solves problem \EnumInc$(P,\Sigma,\pi,\cX)$, for any $\cX\subseteq\cC_{\pi}(\Sigma)$, in time $\poly(|P|,|\Sigma|,|\pi|,|\cX|)$. $\cA$ is said to be a {\it polynomial-delay} algorithm\footnote{Note that checking, for a given $X\in\cC_{\pi}(\Sigma)$, whether $X\in\cX$ can be done in $O(\log|\cX|)=O(|P|)$, by maintaining a priority queue  on the elements of $\cX$, using some lexicographic ordering on the subsets of $P$, see \cite{JYP88}.} for \Enum$(P,\Sigma,\pi)$ if it solves \EnumInc$(P,\Sigma,\pi,\cX)$ in time $\poly(|P|,|\Sigma|,|\pi|)$, independent of $|\cX|$.

When $\pi(X)=1$ for all $X\in\cC(\Sigma)$ and we drop the minimality requirement, \Enum$(P,\Sigma,\pi)$ reduces to enumerating all elements of $\cC(\Sigma)$, a problem known to be solvable with polynomial delay (see, e.g., \cite{BDVG18,NR99} and the references therein). In this section, we consider the case when the property $\pi$ is given by a monotone system of inequalities:
\begin{align}\label{pi}
\pi(X)=1 \quad \Longleftrightarrow\quad f_i(X)\ge t_i, ~\text{ for } i=1,\ldots,m,
\end{align}
where $f_1,\ldots,f_m:2^P\to\RR_+$ are given non-negative {\it monotone} functions, that is, $f_i(X)\le f_i(Y)$ whenever $X\subseteq Y\subseteq\cC(\Sigma)$. We are particularly interested in the class of monotone {\it transversal} functions  $f:2^P\to\RR_+$, defined as follows. Given a hypergraph $\cH\subseteq 2^P$ and non-negative weights $w:\cH\to\RR_+$, $f_{\cH}(X):=\sum_{H\in \cH:~H\cap X\neq\emptyset}w(H)$.	

This class of functions can be  motivated by the following examples:

\smallskip

\noindent {\it Example 1}: Given non-negative numbers $t_1,\ldots,t_m$, and (linear) weight functions $w_1,\ldots,w_m:P\to\RR_+$ on the attribute set, enumerate all closed sets w.r.t. $\Sigma$ whose $i$th weight is at least $t_i$, for all $i=1,\ldots,m$. This can be written in the form \raf{pi} with $m$ linear inequality $f_i(X)=w_i(X):=\sum_{x\in X}w_i(x)\ge t_i$.\footnote{In the case of a single inequality of unit weights, $w\equiv1$, a simple polynomial-delay algorithm can used to enumerate the required closed sets, based on exchanging two elements, using the fact that for every set $X\in\cC(\Sigma)$ there is a set $X'\in\cC(\Sigma)$ such that $|X'|=|X|-1$ (and thus the lattice $\cC(\Sigma)$ is {\it ranked}). When $P$ is an antichain and the weights are arbitrary, a polynomial-delay algorithm also exists based on ordering the elements of $P$ according to the weights, and noting that, for any minimal set satisfying  $w(X)\ge t$, if we delete an element $x\in X$, then we can ``restore feasibility" by adding only one other element $y\in P\setminus X$ such that $w(y)\ge w(x)$ (see, e.g., \cite{LLK80}). When $P$ is a general poset, the same approach does not seem to work (at least immediately), as deleting one  element $x\in X$ may enforce deleting more elements (all elements in $x^+$ have to be deleted); thus restoring feasibility may require adding more than one element. It is plausible that a variant of this approach (e.g., based on deleting only from maximal elements in $X$ w.r.t. the order on $P$) can work for the case of one inequality, but investigating this further is out of the scope of the current paper.}

\smallskip

\noindent {\it Example 2}: 
Given an integer $t$, enumerate all minimal closed sets w.r.t. $\Sigma$ that contain items from at least $t$ rows in $\Sigma$. This can be written in the form \raf{pi} with one transversal inequality $f(X):=|\{D\in\cD:~D\cap X\neq\emptyset\}|\ge t$. 

\smallskip

\begin{figure}
	\centering
	\includegraphics[width=5in]{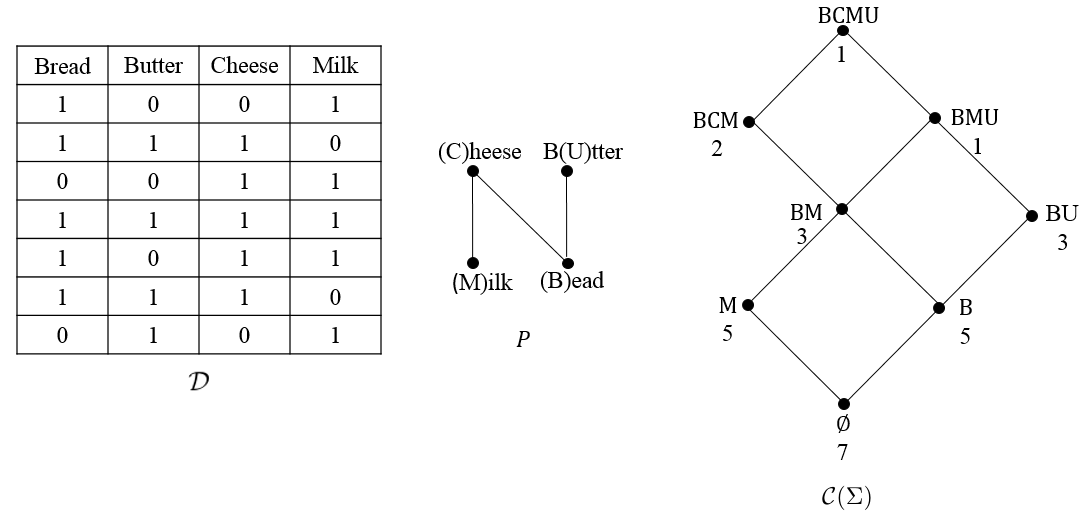}
	\caption{A relational database $\cD$ (supermarket data) and the poset $P:=\{\text{Bread, Butter, Cheese, Milk}\}$ defined by the implication base $\Sigma:=\big\{(\{\text{Butter}\}\rightarrow\text{Bread}),(\{\text{Cheese}\}\rightarrow\text{Bread}),(\{\text{Cheese}\}\to\text{Milk})\big\}$ (for instance, the first implication can be interpreted as ``customers purchasing Butter tend also purchase Bread". The lattice of closed sets w.r.t. $\Sigma$ is shown on the right and the size of the support of each element is shown below the element; for instance, $\supp_{\cD}(\{\text{Bread, Cheese, Milk}\})=2$. For $t=3$, the family of minimal $t$-infrequent closest sets is $\big\{\{\text{Bread, Butter}\},\{\text{Bread, Milk}\}\big\}$.  
	}
	\label{f2}
\end{figure}

\noindent {\it Example 3}: Given an integer $t$, enumerate all minimal closed sets $X$ w.r.t. $\Sigma$ that are contained in at most $t$ rows in $\cD$: $|\supp_\cD(X)|\le t$. These are called {\it minimal $t$-infrequent closed sets}, and
can be expressed as the minimal solutions of~\raf{pi} with one transversal inequality $f(X):=|\{D\in\cD:~\overline D\cap X\neq\emptyset\}|\ge |\cD|-t$. See Figure~\ref{f2} for an example.

\medskip

To a family $\cX\subseteq\cC(\Sigma)$, we associate the (dual) family $\cX^d$ consisting of all (inclusion-wise) maximal sets in $\cC(\Sigma)$ that do {\it not} contain any set in $\cX$.
In particular, $\cC_{\pi}^d(\Sigma)$ is the family of maximal sets in $\cC(\Sigma)$ that do not satisfy property $\pi$. Note that, given any set $X\in\cC(\Sigma)$ such that  $\pi(X)=1$ (resp., $\pi(X)=0$), we can find a set $X'\in\cC_{\pi}(\Sigma)$ (resp., $X'\in\cC^d_{\pi}(\Sigma)$) such that $X'\subseteq X$ (resp., $X'\supseteq X$), in $O(|X|)$ (resp., $O(|P\setminus X|)$) evaluations of $\pi$ by the greedy approach: Initialize $X'=X$, and for an arbitrary order of the elements of $X:=\{x_1,\ldots,x_k\}$ (resp., $P\setminus X:=\{p_1,\ldots,p_k\}$), if $\pi(X'\setminus x_i^+)=1$ set $X'=X'\setminus x_i^+$ (resp., if $\pi(X'\cup p_i^-)=0$ set $X'=X'\cup x_i^-$ ). Let us use $\min_\pi(X)$ (resp., $\max_\pi(X)$) to denote the set computed by this procedure for a given set $X\in\cC(\Sigma)$.

\medskip

Following \cite{BGKM01}, we consider the following {\it joint-generation} problem:

\begin{itemize}
	\item[] \EnumJoinInc$(P,\Sigma,\pi,\cX,\cY)$: Given an implication base $\Sigma$ over a  set $P$, a monotone property $\pi:2^P\to\{0,1\}$, and sets $\cX\subseteq \cC_{\pi}(\Sigma)$ and $\cY\subseteq \cC_{\pi}^d(\Sigma)$, either find a set $Z\in(\cC_{\pi}(\Sigma)\cup\cC_{\pi}^d(\Sigma))\setminus(\cX\cup\cY)$, or declare that no such set exists. 
\end{itemize}

The following is a straightforward extension of the results in \cite{GK99,BI95}, which were proved for the case when $L$ is a Boolean lattice:

\begin{proposition}\label{p6}
	Problems  \LDL$(L,\cA,\cB)$ and \EnumJoinInc$(P,\Sigma,\pi,\cX,\cY)$, where  $L=\cC(\Sigma)$ is given by an implication base $\Sigma$ over a set $P$,  are polynomially equivalent.
\end{proposition}
\proof
Given an instance of problem  \LDL$(L,\cA,\cB)$, with $L=\cC(\Sigma)$ for a given $\Sigma$, which we would like to solve, we construct an instance of problem \EnumJoinInc$(P,\Sigma,\pi,\cX,\cY)$ as follows. Define 
$\pi:2^P\to\{0,1\}$ as either $\pi=\pi_1$ or $\pi=\pi_2$, where
\begin{align*}
\pi_1(X)&=1 ~\text{ if and only if  $X\cap \overline B\neq\emptyset$ for all $B\in\cB$, and} \\
\pi_2(X)&=1 ~\text{ if and only if  $|\{B\in\cB:~X\cap \overline B\neq\emptyset\}|\ge |\cB|$.} 
\end{align*}
In the former case, $\pi$ is defined by a monotone system of linear inequalities, and in the latter case, $\pi$ is defined by one transversal inequality. Define further $\cX:=\{\min_\pi(A):~A\in\cA\}$ and $\cY:=\{\max_\pi(B):~B\in\cB\}$.  Note that $\cX$ is well-defined by~\raf{dual-cond1}. If there is an $X\in\cX$ such that $X\not\supseteq A$ for all $A\in\cA$, then $X$ satisfies \raf{witness1} and thus $(\cA,\cB)$ is declared to be not dual. 
Thus we may assume that $\cX\subseteq\cA$. 
We claim, under this assumption, that $\cX=\cC_{\pi}(\Sigma)$ and $\cY=\cC^d_{\pi}(\Sigma)$ if and only if $(\cA,\cB)$ is a dual pair in $L$. To see this claim, assume first that $(\cA,\cB)$ is dual in $L$. Suppose that there is an $X\in\cC_{\pi}(\Sigma)\cup\cC_{\pi}^d(\Sigma))\setminus(\cX\cup\cY)$. If $X\in\cC_{\pi}(\Sigma)\setminus\cX$, then $X\cap\overline B\neq\emptyset $ for all $B\in\cB$ (as $\pi(X)=1$) and $X\not\supseteq A$ for all $A\in\cA$ (as otherwise $X\supseteq A\supseteq\min_\pi(A)\in\cX\subseteq\cC_{\pi}(\Sigma)$ would imply that $X=\min_\pi(A)\in\cX$).  Thus, $X$ satisfies \raf{witness1} w.r.t. the pair $(\cA,\cB)$, in contradiction to its assumed duality. If $X\in\cC^d_{\pi}(\Sigma)\setminus\cY$, then  $X\subseteq B\subseteq\max_{\pi}(B)\in\cY\subseteq\cC_{\pi}^d(\Sigma)$ for some $B\in\cB$ (as $\pi(X)=0$) would already give the contradiction that $X=\max_{\pi}(B)\in\cY$. Assume next that $\cX=\cC_{\pi}(\Sigma)$ and $\cY=\cC^d_{\pi}(\Sigma)$. If there is an $X$ satisfying \raf{witness1} w.r.t. the pair $(\cA,\cB)$, then $\pi(X)=1$ and $X\not\supseteq A$ for all $A\in\cA$. By our assumption that $\cX\subseteq\cA$, we get $\min_{\pi}(X)\in\cC_{\pi}(\Sigma)\setminus\cX$, a contradiction.

\medskip

Conversely, given an instance of problem \EnumJoinInc$(P,\Sigma,\pi,\cX,\cY)$, for some monotone property $\pi$, which we would like to solve, we construct an instance of problem \LDL$(L,\cA,\cB)$ as follows. We set $L:=\cC(\Sigma)$, $\cA:=\cX$ and $\cB:=\cY$. Clearly, \raf{dual-cond1} holds by monotonicity of $\pi$ as $\pi(A)=1$ for all $A\in\cA$ and $\pi(B)=0$ for all $B\in\cB$. We claim that $\cX=\cC_{\pi}(\Sigma)$ and $\cY=\cC^d_{\pi}(\Sigma)$ if and only if $(\cA,\cB)$ is a dual pair in $L$. To see this claim, assume first that $(\cA,\cB)$ is dual in $L$. Suppose that there is an $X\in\cC_{\pi}(\Sigma)\cup\cC_{\pi}^d(\Sigma))\setminus(\cX\cup\cY)$. If $X\in\cC_{\pi}(\Sigma)\setminus\cX$, then $X\cap \overline B\ne\emptyset$ for all $B\in\cB$ (as $\pi(X)=1$), and $X\not\supseteq A$ for all $A\in\cA$ (as $X\not\in\cX\subseteq\cC_{\pi}(\Sigma)$).  If $X\in\cC^d_{\pi}(\Sigma)\setminus\cY$, then  $X\cap \overline B\neq\emptyset$ for all $B\in\cB$ (as $X\not\in\cY$), and $X\not\supseteq A$ for all $A\in\cA$ (as $\pi(X)=0$). In both cases, $X$ satisfies \raf{witness1} w.r.t. the pair $(\cA,\cB)$, in contradiction to its assumed duality. Assume next that $\cX=\cC_{\pi}(\Sigma)$ and $\cY=\cC^d_{\pi}(\Sigma)$. If there is an $X$ satisfying \raf{witness1} w.r.t. the pair $(\cA,\cB)$, then $X\not\subseteq B$ for all $B\in\cY$ and $X\not\supseteq A$ for all $A\in\cX$. If $\pi(X)=1$, we get that $\min_\pi(X)\in\cC_\pi(\Sigma)\setminus \cX$, and if  $\pi(X)=0$, we get that $\max_\pi(X)\in\cC^d_\pi(\Sigma)\setminus \cY$; in both cases, we get a contradiction.
\qed

It was shown in \cite{BGKM04} that, when $P$ is an antichain, the family $\cC_{\pi}(\Sigma)$ is  {\it uniformly dual-bounded} in the sense defined by the following theorem, for which we state the immediate generalization for any poset $P$ (we only need to observe, by the lattice property, that the family $\cC(\Sigma)$ is closed under unions (and intersections)):

\begin{theorem}[\cite{BGKM01}]\label{t-udbdd}
	Let $f_1, \ldots, f_m: 2^P \to \RR_+$ be $m$ transversal functions defined by $m$ hypergraphs $\cH_1,\ldots, \cH_m$. Let $\cX\subseteq \cC_{\pi}(\Sigma)$ be an arbitrary non-empty subset of the family of minimal closed sets w.r.t. $\Sigma$ satisfying~\raf{pi}. Then  
	\begin{align}\label{udbdd}
	|\cX^d\cap\cC^d_{\pi}(\Sigma)|\le\sum_{i=1}^m|\cH_i|\cdot|\cX|.
	\end{align}
\end{theorem} 

We obtain the following corollary from theorems~\ref{t-main},~\ref{t-udbdd} and Proposition~\ref{p6}, which is a generalization of the result proved for $\Sigma=\emptyset$  ($\dim(\Sigma)=-\infty$) in \cite{BGKM03}.
\begin{corollary}[see \cite{BGKM01}]\label{c3}
	When $\pi$ is defined by a monotone system of transversal inequalities and $\dim(\Sigma)=1$, problem \EnumInc$(P,\Sigma,\pi,\cX)$ can be solved in quasi-polynomial time .
\end{corollary}
\proof
Given $\cX$, we initialize $\cY=\emptyset$ and solve at most $\sum_{i=1}^m|\cH_i|\cdot|\cX|+1$ instances of \EnumJoinInc$(P,\Sigma,\pi,\cX,\cY)$, each time extending either $\cX$ or $\cY$ by a set $Z\in(\cC_{\pi}(\Sigma)\cup\cC_{\pi}^d(\Sigma))\setminus(\cX\cup\cY)$. If $Z\in\cC_{\pi}(\Sigma)\setminus\cX$, we obtain a new minimal closed set satisfying $\pi$. If $Z\in\cC_{\pi}^d(\Sigma)\setminus \cY$ and $Z\not\in\cX^d$, then $Z\not\supseteq X$ for all $X\in\cX$  (as $\pi(Z)=0$ while $\pi(X)=1$ for all $X\in\cX$), and there must exist a set $Y\in\cX^d$ such that $Y\supset Z$ and $\pi(Y)=1$ (as otherwise, $\pi(Y\cup p^-)=1$, for all $p\in\overline Y$, would imply that $Y\in\cC_\pi^d(\Sigma)$), and hence $\min_\pi(Y)\in\cC_{\pi}(\Sigma)\setminus\cX$. In all other cases, $Z\in\cX^d\cap \cC_{\pi}^d$, and it is discarded. By Theorem~\ref{t-udbdd}, the number of discarded sets is not more than $\sum_{i=1}^m|\cH_i|\cdot|\cX|$, and hence, if no new $Z\in\cC_{\pi}\setminus\cX$ is found in $\sum_{i=1}^m|\cH_i|\cdot|\cX|+1$ iterations, we must have $\cC_{\pi}(\Sigma)=\cX$.  
\qed

In contrast to the positive result in Corollary~\ref{c3},  it was shown in \cite{DN20} that, if $L=\cC(\Sigma)$ is given as the lattice of closed sets for an implication base with $\dim(\Sigma)\ge 2$, then problem \LDL$(L,\cA,\cB)$ is generally NP-hard. By an argument similar to the one given in the proof of Proposition~\ref{p6}, we obtain the following result\footnote{Note that when $\dim(\Sigma)=1$, $\cC_{\pi}(\Sigma)$ consists of only one set and the enumeration problem over $\cC(\Sigma)$ is trivial. Thus, the complexity of enumerating minimal $t$-infrequent closed sets. w.r.t. an implication base $\Sigma$ is well-understood for all values of $\dim(\Sigma)$.}.

\begin{theorem}\label{t3}
	Minimal $t$-infrequent closed sets. w.r.t. an implication base $\Sigma$ and a database $\cD$, cannot be enumerated in output polynomial time, unless P=NP, even when $\dim(\Sigma)=2$. 
\end{theorem}
\proof
We start with the  construction in~\cite{DN20} of an NP-hard instance of problem \LDL$(\cC(\Sigma),\cA,\cB))$ over a set $P$ of attributes, where $\dim(\Sigma)=2$. Define $\cD:=\cB$, $\cX:=\{\min_\pi(A):~A\in\cA\}$, where $\pi:2^P\to\{0,1\}$ is defined by: $\pi(X)=1$ if and only if $|\supp_{\cD}(X)|=0$. It is important to note in this construction that $\cX=\cA$ (in fact, $|\cA|=1$). Then $\cC_{\pi}(\Sigma)$ is the  family of $0$-infrequent closed sets, and it is enough to observe (by a similar argument as the one used in the proof of Proposition~\ref{p6}) that the pair $(\cA,\cB)$ is dual in $L:=\cC(\Sigma)$ if and only if $\cC_{\pi}(\Sigma)=\cX$.
\qed

We conclude with the following related result. Given a database $\cD$, an implication base $\Sigma$, and and an integer $t$, a {\it maximal $t$-frequent closed set}, is a set $X\in\cC(\Sigma)$ maximal with the property that $|\supp_\cD(X)|\ge t$. While it seems more natural to consider the family of maximal $t$-frequent sets, it was shown in \cite{BGKM03} that it cannot be enumerated in output polynomial time, unless P=NP, even when $L=\cC(\Sigma)$ is the Boolean lattice.

\nocite{*}
\bibliographystyle{abbrvnat}

\begin{thebibliography}{18}
	\providecommand{\natexlab}[1]{#1}
	\providecommand{\url}[1]{\texttt{#1}}
	\expandafter\ifx\csname urlstyle\endcsname\relax
	\providecommand{\doi}[1]{doi: #1}\else
	\providecommand{\doi}{doi: \begingroup \urlstyle{rm}\Url}\fi
	
	\bibitem[Babin and Kuznetsov(2017)]{BK17}
	M.~A. Babin and S.~O. Kuznetsov.
	\newblock Dualization in lattices given by ordered sets of irreducibles.
	\newblock \emph{Theoretical Computer Science}, 658:\penalty0 316 -- 326, 2017.
	
	\bibitem[Bertet et~al.(2018)Bertet, Demko, Viaud, and Gu\'{e}rin]{BDVG18}
	K.~Bertet, C.~Demko, J.-F. Viaud, and C.~Gu\'{e}rin.
	\newblock Lattices, closures systems and implication bases: A survey of
	structural aspects and algorithms.
	\newblock \emph{Theoretical Computer Science}, 743:\penalty0 93 -- 109, 2018.
	
	\bibitem[Bioch and Ibaraki(1995)]{BI95}
	J.~C. Bioch and T.~Ibaraki.
	\newblock Complexity of identification and dualization of positive boolean
	functions.
	\newblock \emph{Information and Computation}, 123\penalty0 (1):\penalty0
	50--63, 1995.
	
	\bibitem[Boros et~al.(2001)Boros, Gurvich, Khachiyan, and Makino]{BGKM01}
	E.~Boros, V.~Gurvich, L.~Khachiyan, and K.~Makino.
	\newblock Dual-bounded generating problems: Partial and multiple transversals
	of a hypergraph.
	\newblock \emph{SIAM Journal on Computing}, 30\penalty0 (6):\penalty0
	2036--2050, 2001.
	
	\bibitem[Boros et~al.(2002)Boros, Elbassioni, Gurvich, Khachiyan, and
	Makino]{BEGKM02-SICOMP}
	E.~Boros, K.~Elbassioni, V.~Gurvich, L.~Khachiyan, and K.~Makino.
	\newblock Dual-bounded generating problems: All minimal integer solutions for a
	monotone system of linear inequalities.
	\newblock \emph{SIAM Journal on Computing}, 31\penalty0 (5):\penalty0
	1624--1643, 2002.
	
	\bibitem[Boros et~al.(2003)Boros, Gurvich, Khachiyan, and Makino]{BGKM03}
	E.~Boros, V.~Gurvich, L.~Khachiyan, and K.~Makino.
	\newblock On maximal frequent and minimal infrequent sets in binary matrices.
	\newblock \emph{Ann. Math. Artif. Intell.}, 39\penalty0 (3):\penalty0 211--221,
	2003.
	
	\bibitem[Boros et~al.(2004)Boros, Gurvich, Khachiyan, and Makino]{BGKM04}
	E.~Boros, V.~Gurvich, L.~Khachiyan, and K.~Makino.
	\newblock Dual-bounded generating problems: weighted transversals of a
	hypergraph.
	\newblock \emph{Discret. Appl. Math.}, 142\penalty0 (1-3):\penalty0 1--15,
	2004.
	
	\bibitem[Defrain and Nourine(2020)]{DN20}
	O.~Defrain and L.~Nourine.
	\newblock Dualization in lattices given by implicational bases.
	\newblock \emph{Theoretical Computer Science}, 814:\penalty0 169 -- 176, 2020.
	
	\bibitem[Defrain et~al.(2021)Defrain, Nourine, and Uno]{DNU19}
	O.~Defrain, L.~Nourine, and T.~Uno.
	\newblock On the dualization in distributive lattices and related problems.
	\newblock \emph{Discrete Applied Mathematics}, 300:\penalty0 85--96, 2021.
	
	\bibitem[Elbassioni(2008)]{E08}
	K.~Elbassioni.
	\newblock On the complexity of monotone dualization and generating minimal
	hypergraph transversals.
	\newblock \emph{Discrete Applied Mathematics}, 156\penalty0 (11):\penalty0
	2109--2123, 2008.
	
	\bibitem[Elbassioni(2009)]{E09}
	K.~Elbassioni.
	\newblock Algorithms for dualization over products of partially ordered sets.
	\newblock \emph{{SIAM} J. Discret. Math.}, 23\penalty0 (1):\penalty0 487--510,
	2009.
	
	\bibitem[Fredman and Khachiyan(1996)]{FK96}
	M.~L. Fredman and L.~Khachiyan.
	\newblock On the complexity of dualization of monotone disjunctive normal
	forms.
	\newblock \emph{Journal of Algorithms}, 21:\penalty0 618--628, 1996.
	
	\bibitem[Gurvich and Khachiyan(1999)]{GK99}
	V.~Gurvich and L.~Khachiyan.
	\newblock On generating the irredundant conjunctive and disjunctive normal
	forms of monotone \text{Boolean} functions.
	\newblock \emph{Discrete Applied Mathematics}, 96-97\penalty0 (1):\penalty0
	363--373, 1999.
	
	\bibitem[Johnson et~al.(1988)Johnson, Papadimitriou, and Yannakakis]{JYP88}
	D.~S. Johnson, C.~H. Papadimitriou, and M.~Yannakakis.
	\newblock On generating all maximal independent sets.
	\newblock \emph{Information Processing Letters}, 27\penalty0 (3):\penalty0
	119--123, 1988.
	
	\bibitem[Lawler et~al.(1980)Lawler, Lenstra, and Kan]{LLK80}
	E.~Lawler, J.~K. Lenstra, and A.~H.~G.~R. Kan.
	\newblock Generating all maximal independent sets: {NP}-hardness and
	polynomial-time algorithms.
	\newblock \emph{SIAM Journal on Computing}, 9:\penalty0 558--565, 1980.
	
	\bibitem[Nourine and Raynaud(1999)]{NR99}
	L.~Nourine and O.~Raynaud.
	\newblock A fast algorithm for building lattices.
	\newblock \emph{Information Processing Letters}, 71\penalty0 (5):\penalty0 199
	-- 204, 1999.
	
	\bibitem[Stanley and Fomin(1999)]{SF99}
	R.~P. Stanley and S.~Fomin.
	\newblock \emph{Enumerative Combinatorics}, volume~2 of \emph{Cambridge Studies
		in Advanced Mathematics}.
	\newblock Cambridge University Press, 1999.
	
	\bibitem[Wild(2017)]{W17}
	M.~Wild.
	\newblock The joy of implications, aka pure horn formulas: Mainly a survey.
	\newblock \emph{Theoretical Computer Science}, 658:\penalty0 264 -- 292, 2017.
	
\end{thebibliography}
\label{sec:biblio}

\end{document}